\documentclass[a4paper,12pt]{article}
\usepackage[pctex32]{graphics}

\begin{document}
\topmargin 0pt \oddsidemargin 0mm
\newcommand{\beq}{\begin{equation}}
\newcommand{\eeq}{\end{equation}}
\newcommand{\beqa}{\begin{eqnarray}}
\newcommand{\eeqa}{\end{eqnarray}}
\newcommand{\sr}{\sqrt}
\newcommand{\fr}{\frac}
\newcommand{\mn}{\mu \nu}
\newcommand{\G}{\Gamma}

\begin{titlepage}

\vspace{5mm}
\begin{center}
{\Large \bf Logarithmic corrections to entropy for  black holes
with hyperbolic horizon} \vspace{12mm}

{\large   Yun Soo Myung \footnote{e-mail
 address: ysmyung@inje.ac.kr}}
 \\
\vspace{10mm} {\em Institute of Basic Science and School of
Computer Aided Science \\ Inje University, Gimhae 621-749, Korea}
\end{center}
\vspace{5mm} \centerline{{\bf{Abstract}}}
 \vspace{5mm}
We compute logarithmic corrections to entropy for black holes with
hyperbolic horizon. For this purpose, we introduce the topological
black hole and MTZ black holes in four dimensions,  while in  five
dimensions, the topological black holes and Gauss-Bonnet black
hole with negative coupling are considered. As they stand,
logarithmic corrections are problematic  because small black holes
with positive heat capacity have
 negative energies. However, introducing the background state of
extremal black hole, the logarithmic corrections are performed for
black hole with hyperbolic horizon.
\end{titlepage}
\newpage
\renewcommand{\thefootnote}{\arabic{footnote}}
\setcounter{footnote}{0} \setcounter{page}{2}

\section{Introduction}
A number of authors have  shown that for a large class of black
holes, the Bekenstein-Hawking entropy receives logarithmic
corrections due to thermodynamic fluctuations around thermal
equilibrium~\cite{KM1,DMB}. Up to now, a corrected-entropy formula
takes the form
 \beq \label{CEN} S_c=S-\fr{1}{2} \ln \Big[CT^2\Big] + \cdots, \eeq
where $C$ and $T$ are the  heat capacity and temperature of the
given black hole, and $S$ denotes the uncorrected
Bekenstein-Hawking entropy. Here $C$ should be positive for
Eq.(\ref{CEN}) to be well-defined. We note that for $C>0~(C<0)$,
the system is thermodynamically stable (unstable).  A black hole
with negative specific heat is in an unstable equilibrium with the
heat reservoir of the temperature $T$ \cite{GPY}. Its fate under
small fluctuations will be either to decay to  hot flat space or
to grow without limit by emitting or absorbing thermal radiation
in the heat reservoir~\cite{York}. Hence it is meaningless to
apply Eq.(\ref{CEN}) to the black hole with $C<0$. There exists a
way to achieve a stable black hole in an equilibrium with the heat
reservoir. A black hole could be rendered thermodynamically stable
by placing it in AdS spacetimes. An important point is to
understand how a black hole with positive specific heat could
emerge from thermal radiation through a phase transition. To this
end, one introduces the Hawking-Page phase transition between
thermal AdS space and Schwarzschild-AdS black
hole~\cite{HP,BCM,Witt}.

It is well known that the entropy-corrected formula of
Eq.(\ref{CEN}) is universal~\cite{DMB}, implying that this
prescription could apply to all black holes with positive specific
heat.  We note that  the canonical ensemble was used to derive
Eq.(\ref{CEN}). In this case,  the positive energy (mass) of the
system is needed  to define the canonical ensemble. However, small
black holes with hyperbolic horizon have the negative energy, even
though they have positive heat capacity.  These may be  counter
examples, because they have negative masses in the range $M_e \le
M \le 0$ with $M_e$ mass of extremal black hole. Hence, a direct
application of Eq.(\ref{CEN}) to these black holes is not
guaranteed for accounting their thermodynamic fluctuations.

 In this work, we address this issue and resolve this by introducing the background state of
 extremal black hole, similar to the charged black holes with spherical horizon~\cite{CEJM}.
The difference is that for hyperbolic horizon, its background
state energy is negative, whereas its background state energy is
positive for spherical horizon. In order to avoid negative mass,
we use the substraction scheme.  The entropy-corrected formula  is
not changed  because the substraction scheme corresponds to a
constant shifting from $M$ to $E=M-M_e$. We will use
Eq.(\ref{CEN}) to study thermal fluctuations of  black holes with
hyperbolic horizon.

 The organization of this work is as follows. Section 2 is
devoted to performing the logarithmic corrections to topological
black hole (4DTBH) and MTZ black holes in four dimensions. We
study the logarithmic corrections to topological black hole
(5DTBH) and Gauss-Bonnet black holes in five dimensions in section
3. We discuss the AdS/CFT correspondences for these black hole in
section 4. Finally, we discuss our results in section 5.

\section{4DTBH and MTZ black holes}
Topological black holes  in asymptotically anti-de Sitter
spacetimes were first found in three and four
dimensions~\cite{Lemos}. Their black hole horizons are Einstein
spaces of spherical ($k=1$), hyperbolic ($k=-1$), and flat ($k=0$)
curvature for higher dimensions more than three~\cite{Vanzo,BLP}.
The standard equilibrium and off-equilibrium thermodynamic
analyses are possible to show that they are treated as the
extended thermodynamic systems,  even though their horizons are
not spherical.
 The topological
black holes  in four-dimensional AdS spacetimes are given by
 \beq ds^{2}_{4DTBH}=g_{\mu\nu}dx^{\mu}dx^{\nu}=-f_T(r)dt^2 +\fr{1}{f_T(r)}dr^2+r^2d\Sigma_k^2, \label{4DBMT} \eeq
 where the metric function $f_T(r)$ is given by
\beq f_T(r)=k-\fr{m}{r}+ \fr{ r^2}{l^2}. \eeq  $d\Sigma_k^2$
describes the 2D horizon geometry with a constant curvature \beq
d\Sigma_k^2= d\theta^2+ f^2_k(\theta) d\phi^2, \eeq where
$f_k(\theta)$ is given by \beq  f_{0}(\theta) =\theta,
~f_{1}(\theta) =\sin \theta, ~f_{-1}(\theta) =\sinh \theta. \eeq
Here we define $k$=1,~0, and $-1$ cases as the 4D
Schwarzschild-AdS black hole (4DSAdS), 4D flat-AdS black hole, and
4D hyperbolic-AdS black hole (=4DTBH)~\cite{myungsds},
respectively. In the case of $k=1,m=0$, we have an
AdS$_4$-spacetime with its curvature radius $l$. However, $m
\not=0$ generates the topological  black holes. It is easy to
check that the metric (\ref{4DBMT}) satisfies Einstein's equations
with a negative cosmological constant \beq
R_{\mu\nu}=-\frac{3}{l^2} g_{\mu\nu}, \eeq when the horizon is an
Einstein space \beq R_{ij}=k h_{ij}. \eeq In this work, we are
interested in the negative curvature with $k=-1$ only. Then, the
horizon space is a hyperbolic manifold of
$\Sigma_{k=-1}=H^2/\Gamma$, where $H^2$ is 2D hyperbolic space and
$\Gamma$ is a suitable discrete subgroup of the isometry group of
$H^2$~\cite{BM}.

First of all, the 4DTBH provides thermodynamic quantities of
Hawking temperature $T_T$, mass $M_T$, entropy $S_T$, heat
capacity $C_T$, and free energy $F_T=M_T-T_TS_T$
 \beqa \label{TBH}
T_T&=&\frac{1}{4\pi
\rho_+}\Big(\frac{3\rho_+^2}{l^2}-1\Big),~M_T\equiv
\frac{\sigma}{8\pi G_4}m=\frac{\sigma
\rho_+}{8\pi G_4 }\Big(\frac{\rho_+^2}{l^2}-1\Big),\\
S_T&=&\frac{\sigma \rho_+^2}{4 G_4 },~C_T=
2\Bigg[\frac{3\rho_+^2-l^2}{3\rho_+^2+l^2}\Bigg]S_T,~
F_T(\rho_+)=-\frac{\sigma \rho_+}{16 \pi G_4
}\Big(\frac{\rho_+^2}{l^2}+1\Big),  \eeqa where $\sigma$ denotes
the area of a unit 2D hyperbolic space $\Sigma_{k=-1}$ and
$\rho_+$ is the outer horizon which satisfies $f_T=0$ with $k=-1$.
We note that the first-law of thermodynamics $dM_T=T_T dS_T$ holds
for the 4DTBH. Also we observe that the entropy satisfies the
area-law.

Importantly,  we observe that the temperature and heat capacity
are positive for $\rho_+>\rho_e=l/\sqrt{3}$, while the mass is
positive only for $\rho_+>l$. This means that any canonical
ensemble is not defined for  $\rho_+<l$. Thus it seems that the
entropy-correction formula of Eq.(\ref{CEN}) is useless for
describing  thermal fluctuations of the small 4DTBH. This arises
mainly because its horizon geometry is hyperbolic.

In order to resolve this problem, we have to choose an appropriate
substraction scheme. In the Reissner-Norstr\"om-AdS black hole,
one has introduced the extremal black hole as the background state
to define an appropriate free energy in the canonical
ensemble~\cite{CEJM}. Similarly, we wish to introduce the extremal
black hole at $\rho_+=\rho_e$ as the background state even though
the charge is absent in the 4DTBH~\cite{BIR}. In this case, we
define the positive energy as \beq E_T=M_T-M_T^e \eeq with respect
to  the negative background energy \beq
M^e_T=M_T|_{\rho_+=\frac{l}{\sqrt{3}}}=-\frac{\sigma l}{12
\sqrt{3} \pi G_4}. \eeq  Although one uses $E_T$ instead of $M_T$,
all thermodynamics except $F_T$ remain unchanged. The free energy
is an important quantity to discuss the global stability and phase
transition.  In this case, it is shifted  by $F^n_M=E_T-T_MS_M$.
Then, for $\rho_+>\rho_e$, we may  use the corrected-entropy
formula to study thermodynamic fluctuations around the equilibrium
4DTBH. We prove it by showing that the root-mean-square energy
fluctuations defined by the square root of \beq <(\Delta
E_T)^2>=C_TT^2_T \eeq remains unchanged under the change from
$M_T$ to $E_T$. In deriving Eq.(\ref{CEN}), one has used
$S_c=S-\frac{1}{2}\ln S''+\cdots$ with $S''=<M^2>-<M>^2=CT^2$. It
is easy to check $<E^2>-<E>^2=<M^2>-<M>^2$ under the  shifting $M
\to E=M-M_e$, which is also confirmed by noting that $CT^2$ is an
invariant quantity. Then, the mass (energy) becomes positive and
thus one could define the canonical ensemble to discuss its
thermal fluctuations around the equilibrium configuration.
Importantly, the entropy-corrected formula is not changed because
the substraction scheme corresponds to a constant shifting from
$M$ to $E$.

For a large 4DTBH with $\rho_+ \gg l$, we have approximate forms
\beq C_T \sim 2S_T,~T_T^2 \sim S_T \eeq which leads to an
approximately correct-entropy formula \beq S^c_T\sim S_T-
\ln\Big[S_T\Big]+\cdots. \eeq  This  is the same form as for large
4DSAdS~\cite{DMB}. Furthermore, we note that the Smarr formula of
$M=TS/2$ (Euler relation) is not satisfied for the choice of
either $M=M_T$ or $E_T$.

On the other hand, the MTZ black hole dressed by scalar could be
obtained from the Einstein action minimally coupled to a scalar in
AdS spacetimes ~\cite{MTZ} \beq I_4[g,\phi]= \int
d^4x\sqrt{-g}\Bigg[\frac{R-2\Lambda_4}{16 \pi
G_4}-\frac{1}{2}(\nabla\phi)^2-V(\phi)\Bigg] \label{4SDS} \eeq
where the potential $V(\phi)$ is given by \beq
V(\phi)=-\frac{3}{4\pi G_4l^2}\sinh^2 \Bigg[\sqrt{\frac{4 \pi
G_4}{3}}\phi\Bigg].\eeq  Here $\Lambda_4=-3/l^2$ with $l$  the
curvature radius of AdS$_4$ spacetimes. For $\phi=0$ case, we
obtain  the action for the 4DTBH. In order to understand the role
of scalar and its potential, it would be better to go the
conformal frame  by introducing  a conformal factor
$\Psi=\sqrt{3/4\pi G_4}\tanh[\sqrt{4\pi G_4/3}\phi]$. Performing a
conformal transformation as $\hat{g}_{\mu\nu}=(1-4\pi
G\Psi^2/3)^{-1}g_{\mu\nu}$, the action $I_4$ takes the from \beq
I_4[\hat{g},\Psi]= \int
d^4x\sqrt{-\hat{g}}\Bigg[\frac{\hat{R}-2\Lambda_4}{16 \pi
G_4}-\frac{1}{2}(\nabla\Psi)^2-\frac{1}{12}\hat{R}\Psi^2-\frac{2\pi
G_4}{3l^2}\Psi^4\Bigg] \label{4SDSc}, \eeq where a conformally
coupled scalar $\Psi$ appears with the conventional potential
$\Psi^4$.  In connection with higher derivative terms, this
potential may play a role of curvature-squared term $\hat{R}^2$.
It provides a black hole dressed by scalar with hyperbolic
horizon. On the other hand, for $\Psi=0$, one has the action for
4DTBH, Eq.(\ref{4DBMT}). Then, the solution of the MTZ black hole
is given by \beq \label{metricf} ds^2_{M}=-f_{M}(r)dt^2+
\frac{dr^2}{f_M(r)}+r^2d\Sigma_{k=-1}^2\eeq where the metric
function $f_M(r)$ is given by  \beq
f_M(r)=\frac{r^2}{l^2}-\Bigg(1+\frac{G_4\mu}{r}\Bigg)^2 \eeq and a
conformally coupled scalar  has the configuration \beq
\bar{\Psi}(r)=\sqrt{\frac{3}{4 \pi G_4}}\frac{G_4\mu}{r+G_4\mu}.
\eeq

Thermodynamic quantities of the MTZ are given by
 Hawking temperature $T_M$, mass $M_M$, entropy
 $S_M$, heat capacity $C_M$, and   free energy $F_M=M_M-T_MS_M$
by~\cite{MTZ,Siop,Myungsh} \beqa \label{MBH}T_M&=&\frac{1}{2\pi
l}\Big[\frac{2r_+}{l}-1\Big]=\Big(\frac{2G_3}{\sigma\pi
l^3}\Big)S_M,~M_M\equiv \frac{\sigma}{4 \pi}\mu=\frac{\sigma
r_+}{4\pi G_4}\Big(\frac{r_+}{l}-1\Big),\\
\nonumber S_M&=&\frac{\sigma l^2}{4
G_4}\Big(\frac{2r_+}{l}-1\Big)=C_M,~ F_M(r_+)=-\frac{\sigma}{8 \pi
G_4}\Big(\frac{2r_+^2}{l}-2r_+ +l\Big), \eeqa where $r_+$  is the
outer horizon which satisfies $f_M=0$.  We note that the first-law
of thermodynamics $dM_M=T_MdS_M$ is satisfied  for the MTZ. Also
we observe that the entropy $S_M$ was obtained by using the Wald's
formula but it  does not satisfy the area-law. Importantly, the
temperature and heat capacity are positive for $r_+>r_e=l/2$,
while the mass is positive for $r_+>l$ only. This means that any
canonical ensemble is not suitable for describing the case of
$r_+<l$ and thus the corrected-entropy formula of Eq.(\ref{CEN})
may be useless for the small MTZ black hole.

In order to resolve this problem, we have to choose an appropriate
background state.  Similarly, we introduce the extremal black hole
as the background state  even though the charge is absent in the
MTZ black hole, too. For the MTZ black hole, we define the
positive energy as~\cite{NVZ} \beq E_M=M_M-M_M^e \eeq with the
background state energy \beq
M^e_M=M_M|_{r_+=\frac{l}{2}}=-\frac{\sigma l}{16 \pi G_4}. \eeq
Although one uses $E_M$ instead of $M_M$, all thermodynamics of
the MTZ except $F^n_M=E_M-T_MS_M$ remain unchanged. Nicely, we
check that the Smarr formula \beq \label{SF} E_M=\frac{T_MS_M}{2}
\eeq is satisfied for $r_+ > r_e$ only when using the positive
energy $E_M$. This implies that  we could use the
corrected-entropy formula to see thermodynamic fluctuations around
the equilibrium MTZ, even though its entropy does not satisfy the
area-law. In this case, we have a corrected-entropy formula for
$r_+>r_e$
 \beq S^c_M= S_M-
\frac{3}{2}\ln\Big[S_M\Big]+\cdots \eeq which is the same formula
for the non-rotating BTZ black hole~\cite{DMB} whose thermodynamic
quantities are given by \beq \label{NBTZ} T_B=\frac{r_+}{2\pi
l^2}=\Big[\frac{G_3}{\pi^2l^2}\Big]S_B,~M_B=\frac{
r_+^2}{8G_3l^2}=E_B,~ S_B=\frac{\pi r_+}{2 G_3}=C_B,~
F_B=-\frac{r_+^2}{8 G_3l^2}.\eeq Here we have the Smarr formula of
 $E_B=T_BS_B/2$ for the non-rotating BTZ black hole. Two are
very similar in the sense that they satisfy the Smarr formula and
have the same corrected-entropy formula. A difference is that
these formulae are valid for the outer horizon $r_+> r_e$ of MTZ
black hole and for any size of non-rotating BTZ black hole.
Furthermore, we derive thermodynamic quantities for large MTZ
black hole with $r_+ \gg l$ as \beq T_M \simeq \frac{r_+}{\pi
l^2},~M_M \simeq \frac{\sigma r_+^2}{4\pi l G_4},~ S_M=C_M\simeq
\frac{\sigma l r_+}{2 G_4},~ F_M \simeq -\frac{\sigma r_+^2}{4 \pi
l G_4}\eeq which show the nearly same behavior as Eq. (\ref{NBTZ})
is shown. Particularly, choosing $\sigma=\pi/l$ leads to close
relations of $T_M=2T_B, M_M=2M_B, F_M=2F_B, S_M=S_B, C_M=C_B$.
This suggests that the MTZ black hole dressed by scalar is a
corner stone for the  4D black hole physics as the non-rotating
BTZ black hole does play  a key role  in the 3D black hole
physics.

\section{5DTBH and Gauss-Bonnet black holes}
In five dimensions, the  topological AdS black holes are  given by
 \beq ds^{2}_{5DTBH}=
 -h(r)dt^2 +\fr{1}{h(r)}dr^2 +r^2d\Sigma^2_k,
\label{BMT} \eeq where $d\Sigma^2_k= d\chi^2
+h^2_{k}(\chi)(d\theta^2+ \sin^2 \theta d\phi^2)$ describes the 3D
horizon geometry with a constant curvature. Further $h(r)$ and
$h_k(\chi)$ are given by \beq h(r)=k-\fr{m}{r^2}+ \fr{
r^2}{\ell^2},~~~ h_{0}(\chi) =\chi, ~h_{1}(\chi) =\sin \chi,
~h_{-1}(\chi) =\sinh \chi. \eeq Here we define $k$=1,~0, and $-1$
cases as the 5D Schwarzschild-AdS black hole (5DSAdS)
~\cite{MP,CMu}, 5D flat-AdS  black hole, and 5D hyperbolic-AdS
black hole (5DTBH)~\cite{CAI1}, respectively.  We are interested
in the $k=-1$ case only. In this case, the location of the  event
horizon is given by
 \beq \label{EH} r_{+}^2=
\fr{\ell^2}{2}\Big(1+\sqrt{1 +4 m/\ell^2}\Big). \eeq The relevant
thermodynamic quantities of Hawking temperature $T_t$, mass $M_t$,
Bekenstein-Hawking entropy $S_t$,   heat capacity $C_t$, and free
energy $F_t=M_t-T_tS_t$ are given by \beqa \label{TQ}
&&T_t=\fr{1}{2 \pi r_+}\Big(\fr{2r^2_{+}}{ \ell^2}-1\Big),~M_t
\equiv \fr{3V_3 m}{16 \pi G_5} =\fr{3V_3 r_+^2}{16 \pi
G_5}\Big(\frac{r_+^2}{l^2}-1\Big),~~\\
\nonumber && S_t=\fr{V_3r_{+}^3}{4G_5},~C_t= 3
\Bigg[\fr{2r_{+}^2-\ell^2}{2r_{+}^2+\ell^2}\Bigg]S_t,~F=-\fr{V_3
r_{+}^2}{16 \pi G_5}\Big(\fr{r_{+}^2}{\ell^2}+1 \Big) \eeqa where
$V_3$ is the volume of a unit 3D hyperbolic space $\Sigma_{k=-1}$
and $G_5$ is the five-dimensional Newton constant. Here we find
that $T_t,~C_t>0$ for $r_+>r_e=l/\sqrt{2}$, while $M_t >0$ for
$r_+>l$. It seems  that any canonical ensemble is not defined for
$r_+<l$ and thus the entropy-correction formula of Eq.(\ref{CEN})
is meaningless for the small 5DTBH.

In order to cure this problem, we have to find an appropriate
substraction scheme. Similarly, we introduce the extremal black
hole as the background state . In this case, we define the
positive energy as \beq E_t=M_t-M_t^e \eeq with the background
state energy \beq M^e_t=M_t|_{r_+=\frac{l}{\sqrt{2}}}=-\frac{3V_3
l^2}{64 \pi G_5}. \eeq Although one uses $E_t$ instead of $M_t$,
all thermodynamics except $F^n_t=E_t-T_tS_t$ remain unchanged.
This implies that for $r_+>r_e$, we could use the
corrected-entropy formula to study thermodynamic fluctuations
around  the equilibrium 5DTBH.

For a large 5DTBH with $r_+ \gg l$, we have approximate forms \beq
C_t \sim 3S_t,~T_t^2 \sim S_t^{\frac{2}{3}} \eeq which leads to an
approximately correct-entropy formula \beq S^c_t\sim S_t-
\frac{5}{6}\ln\Big[S_t\Big]+\cdots. \eeq This  is the same form as
for large 5DSAdS~\cite{DMB,Myu}. Furthermore, we note that the
Smarr formula is not satisfied for the choice of either $M=M_t$ or
$E_t$.

In order to find a 5D version of MTZ black hole, we may consider
the 5D action like Eq.(\ref{4SDSc}) with replacing the last two
terms and $\Lambda_4$ by $(3/32) R \Psi^2+\alpha_5 \Psi^{10/3}$
and $\Lambda_5$. However, it was turned out that there is no
solution like the MTZ black hole in AdS$_5$ spacetimes~\cite{NVZ}.
That is, there is no such a 5D black hole dressed by scalar.
Instead, we consider a similar black hole with hyperbolic horizon.
This may be found from a 5D gravitational action in the presence
of a negative cosmological constant $\Lambda_5 = -6/l^2$ and
Gauss-Bonnet (GB) term as
\begin{equation}\label{baction}
I_5[g,c] = \frac{1}{16\pi G_5} \int \! d^5 x \sqrt{- g}\left[ R
-2\Lambda_5 +\frac{c}{2} L_{GB} \right],
\end{equation}
where
\begin{equation}
L_{GB} =  R^2 -4 R_{\mu\nu} R^{\mu\nu} + R_{\mu\nu\rho\sigma}
R^{\mu\nu\rho\sigma}. \label{GBaction}
\end{equation}
Here $c$ is a GB coupling constant having mass dimension $-2$. For
$c=0$, this action reduces to the action for the 5DTBH,
Eq.(\ref{BMT}).  The solution of the TGBAdS black hole is given by
\begin{equation}\label{met}
  ds^2_g =  - f_g(r) dt^2 +\frac{dr^2}{f_g(r)} + r^2 d\Sigma_{k}^2.
\end{equation}
The metric function  is given by
\begin{equation}
 f_g(r) =  k +\frac{r^2}{2c}\left[1+\epsilon\sqrt{1+\frac{4c}{3}\left(\frac{2\mu
 }{r^4}-\frac{3}{l^2}\right)}\right].
  \label{solf}
\end{equation}
Hereafter, we consider the negative coupling of $c<0$ with $k=-1$
and $\epsilon=-1$  in accordance with the hyperbolic horizon.

For the TGBAdS black hole,  we have  relevant thermodynamic
quantities~\cite{CaiGB,Neup,MKP} \beqa
&&T_g=\frac{r_+}{2\pi(r_+^2-2c)}\left(\frac{2r^2_+}{l^2}-1\right),~~
M_g\equiv \frac{V_3 \mu}{8 \pi G_5}= \frac{3V_3 r_+^2}{16 \pi G_5}
\left(\frac{r^2_+}{l^2}-1+\frac{c}{r_+^2}\right), \\ \nonumber
&&S_g=\frac{V_3 r^3_+ }{4G_5}\left(1-\frac{6c}{r^2_+}\right),~ C_g
=3\frac{ V_3 r_+^3}{4G_5} \Bigg[\frac{ \left(2 r^2_+ - l^2 \right)
\left(1 -2 c/r_+^2 \right)^2 } { (2r^2_+ +  l^2) - (2 c/r_+^2)
\left(6 r^2_+ - l^2\right)}\Bigg], \\ \nonumber
 && F_g=M_g-T_gS_g=-\frac{V_3r_+^4}{16\pi G_5\left(r^2_+-2c\right)}
      \left[\Big(\frac{r^2_+}{l^2}+1\Big)-\frac{3c}{r_+^4}\left(\frac{6r^4_+}{l^2}-r^2_+-2c\right)\right],
\eeqa where we point out that $S_g$ was derived using the
first-law of thermodynamics and thus it does not satisfy the
area-law. Here we find that $T_g,~C_g>0$ for $r_+>r_e=l/\sqrt{2}$.
On the other hand,  one has  $M_t >0$ for $r_+>r_0$ where \beq
r_0=\frac{l}{\sqrt{2}}\sqrt{1+\sqrt{1-4c/l^2}}>l \eeq which is
determined by the massless condition of $M_g=0$. This means that
any canonical ensemble is not defined for $r_+<r_0$ and thus the
entropy-correction formula of Eq.(\ref{CEN}) is useless for the
small TGBAdS.  In order to solve this problem, we have to find an
appropriate background state. We introduce the extremal black hole
as the background state. In this case, we define the positive
energy as \beq E_g=M_g-M_g^e \eeq with the background state
energy\beq M^e_g=M_g|_{r_+=\frac{l}{\sqrt{2}}}=-\frac{3V_3}{16 \pi
G_5}\Bigg[\frac{l^2}{4}-c\Bigg]. \eeq At this stage, we find  an
important equality for energies \beq E_g=E_t=\frac{3V_3r_+^2}{16
\pi G_5}\Bigg[\frac{r_+^2}{l^2}-1+\frac{l^2}{4r_+^2}\Bigg]. \eeq
Although one uses $E_g$ instead of $M_g$, all thermodynamics
except $F^n_g=E_g-T_gS_g$ remain unchanged. This implies that for
$r_+>r_e$, we could use the corrected-entropy formula to see
thermodynamic fluctuations around the equilibrium TGBAdS. For a
large TGBAdS with $r_+ \gg l,c$, we have approximate forms \beq
C_g \sim 3S_g,~T_g^2 \sim S_g^{\frac{2}{3}} \eeq which leads to an
approximately correct-entropy formula \beq S^c_g\sim S_g-
\frac{5}{6}\ln\Big[S_g\Big]+\cdots. \eeq This  is the same form as
for large 5DSAdS~\cite{DMB}. Furthermore, we note that the Smarr
formula is not satisfied for the choice of either $M=M_g$ or
$E_g$.

\section{Boundary conformal field theories}
 The holographic principle means that the number of
degrees of freedom associated with the bulk gravitational dynamics
is determined by its boundary spacetime. We start with the
bulk-relation between the entropy and energy for the non-rotating
BTZ black hole \beq \label{NBTZS}S_B=\pi l
\sqrt{\frac{2E_B}{G_3}}\eeq which is derived using $E_B=\frac{(\pi
lT_B)^2}{2G_3}$ and Smarr formula of $E_B=\frac{T_BS_B}{2}$.

Also this could be recovered from the CFT on the boundary at
infinity using the Cardy formula \beq S_{CFT}=2 \pi
\sqrt{\frac{cL_0}{6}}+2 \pi \sqrt{\frac{\bar{c}\bar{L}_0}{6}}\eeq
with \beq c=\bar{c}=\frac{3l}{2G_3},~L_0=\bar{L}_0=\frac{E_Bl}{2}.
\eeq This means  that the AdS$_3$/CFT$_2$ correspondence was
realized for counting the BTZ black hole entropy.

However, as far as we know, there is no definite realization of
the AdS$_4$/CFT$_3$ correspondence. Hence, we may try to
understand this correspondence by analogy of the non-rotating BTZ
black hole. First of all, we find the similar relation from
$E_M=\frac{\sigma \pi l^3 T_M^2}{4G_4}$ and the Smarr formula of
Eq.(\ref{SF}) as \beq \label{MTZS} S_M=l \sqrt{\frac{\pi l \sigma
E_M}{G_4}}=2\pi l \sqrt{ (-4M_M^e)E_M}\eeq which is valid for the
outer horizon $r_+> r_e$. Assuming that the Cardy formula \beq
\label{card4} S_{CFT}^M=2 \pi l \sqrt{\frac{c_3L^M_0}{6}}+2 \pi l
\sqrt{\frac{\bar{c}_3\bar{L}^M_0}{6}}\eeq with \beq \label{card5}
c_3=\bar{c}_3=\frac{3\sigma l}{2\pi
G_4}=-24M^e_M,~L^M_0=\bar{L}^M_0=\frac{E_M}{2} \eeq holds in three
dimensions, we may recover the entropy $S_M$ from the boundary
CFT$_3$ entropy $S^{M}_{CFT}$.  We note that Eqs.(\ref{card4}) and
(\ref{card5}) are just our proposition. This is because we do not
know what is the exact Cardy formula for  the boundary CFT$_3$.
One thing to show is that one may recover the bulk entropy $S_M$
from the presumed formula (\ref{card4}) with (\ref{card5}).

The AdS$_5$/CFT$_4$ correspondence represents also a concrete
realization of the holographic  principle. In this case, there is
no Smarr formula to derive a direct relation like
Eqs.(\ref{NBTZS}) and (\ref{MTZS}). Instead, one proposes the
Cardy-Verlinde formula
 for a strongly coupled CFT$_4$ with
its AdS dual~\cite{VER}. It is known that this formula holds for
various kinds of asymptotically AdS spacetimes including the TAdS
black holes~\cite{CAI1}.  The boundary spacetimes for the CFT$_4$
are defined through the AdS$_5$/CFT$_4$
correspondence~\cite{witten} \beq \label{bcft}
ds^2_{CFT_4}=\lim_{r \to
\infty}\fr{R^2}{r^2}ds^2_{5DTBH}=-\fr{R^2}{l^2}dt^2 +R^2
d\Sigma^2_k. \eeq
 From the above, the relation between the
five-dimensional bulk and four-dimensional boundary quantities is
given by $E_{CFT}=(l/R)E$ and $ T_{CFT}=(l/R)T$ where the size of
boundary space $R$ satisfies $T_{CFT}>1/R$. As is expected, we
obtain the same entropy: $S_{CFT}=S$. We note that the boundary
system at high temperature is described by the CFT-radiation with
 equation of state $p=E_{CFT}/3V_3$. Then,  the Casimir energy
defined  by $E_c=3(E_{CFT}+pV_3-T_{CFT}S_{CFT})$~\cite{Setare} is
necessary to obtain  the Cardy-Verlinde formula \beq \label{CVF}
S_{CFT}=\frac{2 \pi R}{3\sqrt{|k|}}\sqrt{E_c(2E_{CFT}-E_c)}. \eeq
 The non-zero Casimir energy reflects  that the Euler relation is not satisfied.  We
find the boundary thermal quantities for 5DTBH as functions of
$\hat{r}=r_{+}/l$ ~\cite{myungsds,Myu,myungcqg} \beqa
&&E_{CFT}^{t}=\fr{3V_3\kappa\hat{r}^2}{R}\Bigg[\hat{r}^2-1+\frac{1}{4\hat{r}^2}\Bigg],
~~T_{CFT}^{t}=\fr{\hat{r}}{2\pi
 R}\Big[2\hat{r}^2-1\Big],\\ \nonumber
&&E_c^{t}=-\fr{3V_3 \kappa\hat{r}^2(2\hat{r}^2-1)}{R}\eeqa with
$\kappa=l^3/16 \pi G_5$. On the other hand, for the TGBAdS case,
we find the boundary thermal quantities  \beqa
&&E_{CFT}^{g}=\fr{3V_3\kappa\hat{r}^2}{R}\Bigg[\hat{r}^2-1+\frac{1}{4\hat{r}^2}\Bigg],
~~T_{CFT}^{g}=\fr{\hat{r}}{2\pi
 R(\hat{r}^2-2\hat{c})}\Big[2\hat{r}^2-1\Big],\\ \nonumber
&&E_c^{g}=-\fr{3V_3
\kappa\hat{r}^2(2\hat{r}^2-1)}{R}\Bigg[\frac{(1-8\hat{c})\hat{r}^2-2\hat{c}}{\hat{r}^2-2\hat{c}}\Bigg].
\eeqa

Concerning the AdS/CFT correspondence, we remind the reader that
the boundary CFT$_4$ energy ($E_{CFT}$) should be positive in
order for it to make sense.
 Also the Casimir energy ($E_c$) is related to the central charge
 of the corresponding CFT$_4$. If it is negative, one  may obtain
 a non-unitary CFT$_4$. In this sense, for $\hat{r}>1/\sqrt{2}$, 5DTBH
 provides
 a non-unitary conformal field theory because of
 $E^t_{CFT}>0$ and $E^t_c<0$. Similarly, for TGBAdS with $\hat{c}<0$ and $\hat{r}>1/\sqrt{2}$,
it
 provides
 a non-unitary conformal field theory  with
 $E^g_{CFT}>0$ and $E^g_c<0$. For $\hat{r}<1/\sqrt{2}$, two cases are
 not well defined because of negative energy and  temperature on
 the boundary. Hence we do not have any Cardy-Verlinde formula
 for the 5DTBH and TGBAdS.

Finally, we introduce the Cardy-Verlinde formula (\ref{CVF}) to
show whether or not  the substraction scheme does improve  the
boundary CFT$_4$ of the topological black hole. However, it does
not resolve  an original issue of the non-existence of
Cardy-Verlinde formula for the topological black holes.

\section{Discussion}
We believe that for a large class of black holes, the
Bekenstein-Hawking entropy receives logarithmic corrections due to
thermodynamic fluctuations around thermal equilibrium. However,
small black holes with hyperbolic horizon have the negative
energy, even though they have positive heat capacity. Hence, a
direct application of Eq.(\ref{CEN}) to these black holes is not
suitable for investigating their thermodynamic fluctuations. We
resolve this issue by introducing the mass of extremal black hole
as the background state energy for 4DTBH, MTZ, 5DTBH, TGBAdS black
holes. In order to avoid negative mass problem, we use the
substraction scheme.  The entropy-corrected formula  is not
changed  because the substraction scheme corresponds to a constant
shifting from $M$ to $E=M-M_e$.  We have used Eq.(\ref{CEN}) to
study thermal fluctuations of  black holes with hyperbolic
horizon.

Especially, we find that the MTZ black hole dressed by a scalar
has the best thermodynamic property in AdS$_4$ spacetimes as the
BTZ black hole does show in AdS$_3$ spacetimes. Hence we propose
the MTZ black hole with hyperbolic horizon as the corner stone to
study the 4D black holes in AdS$_4$ spacetimes, better than  the
Schwarzschild-AdS black hole. This is because the
Schwarzschild-AdS black hole with spherical horizon could not
provide a completely thermodynamic picture as  the MTZ black hole
does show~\cite{CMu,Myu}.

\section*{Acknowledgments}
This work was in part  supported by the Korea Research Foundation
Grant (KRF-2005-013-C00018) and the SRC Program of the KOSEF
through the Center for Quantum Spacetime (CQUeST) of Sogang
University with grant number R11-2005-021-03001-0.

\end{document}